\documentclass[12pt]{iopart}
\usepackage{graphicx}
\usepackage{overpic}
\expandafter\let\csname equation*\endcsname=\relax
\expandafter\let\csname endequation*\endcsname=\relax
\usepackage{amsmath,amssymb,amsthm}

\begin{document}

\title[]{Control of laser-plasma instabilities by non-collinear polychromatic light}

\author{Yao Zhao$^{1,5}$, Hongwei Yin$^{1}$, Bin Zhao$^{2,4}$, Zijian Cui$^{3,5}$}

\address{$^1$School of Science, Shenzhen Campus of Sun Yat-sen University, Shenzhen 518107, China}
\address{$^2$School of Mathematics and Physics, Nanjing Institute of Technology, Nanjing 211167, China}
\address{$^3$Key Laboratory on High Power Laser and Physics, Shanghai Institute of Optics and Fine Mechanics, CAS, Shanghai 201800, China}
\address{$^4$Collaborative Innovation Center of IFSA (CICIFSA), Shanghai Jiao Tong University, Shanghai 200240, China}
\address{$^5$Author to whom any correspondence should be addressed.}

\ead{zhaoyao5@mail.sysu.edu.cn,cuizijian@siom.ac.cn}

\begin{abstract}
Normal broadband lasers with collinear polychromatic components have immense potential for mitigating laser plasma instabilities (LPIs). However, the projection complexity of collinear polychromatic light (CPL) is a significant challenge owing to the demand for a large bandwidth and beamlet number. Here, we propose a theoretical LPI model and optical design for non-collinear polychromatic light (NCPL), which has a small angle $\sim4^\circ$ and large frequency difference $\sim$1\% between the double-color beamlets. LPI models of the NCPL demonstrate a decoupling threshold for the shared daughter waves under a multibeam configuration. Compared with the CPL, the wavevector couplings are further reduced by the introduced angle. Therefore, both the growth rate and saturation level of LPIs are greatly reduced by using the NCPL. The two- and three-dimensional simulation results indicate that the NCPL reduces the absolute and convective decoupling thresholds of the CPL and is sufficient to effectively mitigate the reflectivity, hot-electron generation, and intensity of cross-beam energy transfer. An optical design for the efficient generation of ultraviolet NCPL has been presented based on the unsaturated optical parametric amplification and non-collinear sum-frequency generation.

Yao Zhao has corrected a few typos in the manuscript, especially the notations about Eq. (9).
\end{abstract}

\section{Introduction}

After decades of effort \cite{Hinkel2004National,Lindl2014Review,Hurricane2014Fuel,doppner2020achieving}, the ignition of inertial confinement fusion (ICF) was achieved at the National Ignition Facility, which is a new milestone in nuclear fusion energy \cite{abu2022lawson,clery2022explosion}. Despite this, enhancing the beam-target coupling efficiency by controlling laser plasma instabilities (LPIs) remains a critical issue for achieving high-efficiency ignition \cite{froula2010experimental,Pesme2002Laser,ping2019enhanced,tikhonchuk2019studies,Atzeni2021evaluation}. Even though LPIs can be suppressed by low gas-filled hohlraums \cite{hall2017relationship}, the expansion of the hohlraum wall results in the mixing of high-Z ions into the fuel and the degradation of the imploding process \cite{le2020plasma,sio2022fuel,riedel2021kinetic,higginson2022understanding}. Therefore, the mitigation of LPIs without reducing hohlraum performance is critical for robust ignition. LPIs are also inevitable in direct-drive schemes for achieving inertial fusion energy \cite{craxton2015direct,he2016hybrid,zhang2020double}. In addition to their application in ICF, the control of LPIs can suppress plasma heating and wave breaking caused by the intense prepulse and enlarge the applicable parameter range of particle acceleration \cite{batani2010effects,kaluza2004influence} and plasma optics \cite{teubner2009high,mucke2011isolated,pan2023circularly,zhao2020plasma,yang2021wave}.

The mitigation of LPIs using broadband lasers has been investigated since the early 1970s \cite{kruer1973instability,follett2019thresholds,santos2007white,follett2018suppressing,baton2020preliminary,moody2009control,2001Reduction}. The random phase model suggests that the linear growth of LPIs can be reduced if the laser bandwidth is much larger than the growth rate \cite{estabrook1980heating}. The simulation results indicate that an incoherent laser with a small continuous bandwidth always behaves poorly over a long time owing to the coupling of different color components and uncontrollable wave turbulence \cite{zhao2015effects,zhao2017stimulated,brandao,liu2022non}. At present, no practical scheme has been demonstrated for the efficient third-harmonic generation (THG) of broadband lasers \cite{gao2020high,eimerl2016stardriver,dorrer2021broadband,fusaro2021improving} because the phase-matching conditions required for THG are difficult to satisfy \cite{dorrer2022spectral,weaver2017spectral,obenschain2015high}. Collinear polychromatic light (CPL) with multicolor beamlet propagation in the same direction was thus proposed \cite{zhao2021polychromatic,zhao2022mitigation}, which can effectively mitigate LPIs without technical obstacles. However, the optical system is complicated and overloaded for practical applications because each CPL requires at least four color collinear beamlets.

In contrast to conventional broadband lasers \cite{zhao2022mitigation}, in this article, we propose the LPI model and optical design for non-collinear polychromatic light (NCPL) that has a nonzero angle between different-color beamlets with detuning $\gtrsim1\%$. Theoretical models of NCPL show that beamlets are non-collinear if the introduced angle satisfies $\alpha\gtrsim3^\circ$, and therefore the wavevector couplings of LPIs are reduced. Compared to CPL with the same bandwidth, the ponderomotive force and saturation level of convective stimulated Raman scattering (SRS) \cite{guzdar1991effect} with NCPL are reduced by 15\% and more than 24\%, respectively. Analogous to SRS, the intensities of stimulated Brillouin scattering (SBS) \cite{brandao,divol2007controlling}, cross-beam energy transfer (CBET) \cite{dubois1992collective,qiu2021collective} can be further reduced by NCPL. The theoretical model of two-plasmon decay (TPD) instability demonstrates that the compact instability region is dispersed into decoupled submodes with much weak intensity by the NCPL. For symmetry compression, the polychromatic configuration is also used to balance the energy of the inner and outer cones \cite{Hansen2021cross,Oudin2021reduction}. Both two-dimensional (2D) and three-dimensional (3D) particle-in-cell (PIC) simulations indicate that NCPL with double-color beamlets is sufficient to effectively mitigate the reflectivity \cite{montgomery2016two}, hot-electron generation \cite{rosenberg2018origins,christopherson2021direct}, and CBET intensity \cite{marozas2018first,bates2018mitigation} in laser-plasma interactions.

Third-harmonic NCPL with large detuning can be efficiently generated by a series of nonlinear processes, including second-harmonic generation (SHG), unsaturated optical parametric amplification (UOPA), and non-collinear sum-frequency generation (NCSFG). The UOPA and NCSFG processes performed in a single DKDP crystal can significantly reduce the crystal number and result in the same efficiency as that of the traditional driver scheme. Therefore, the NCPL driver with a compact configuration can effectively mitigate LPIs, which paves the way towards the realization of robust and high-efficiency fusion ignition.

\section{Theoretical models of LPIs driven by the non-collinear polychromatic light}

The mitigation conditions of NCPL are studied based on the double-color LPI model, where the incident angle in a vacuum is $\alpha=\alpha_1-\alpha_2=2\alpha_1=-2\alpha_2$ considering spatial symmetry. We first study the dispersion relation of double-color beams interaction with homogeneous plasmas. Without loss of generality, an example for SRS is presented in detail, the fluid equations of which are \cite{kruer1988physics}
\begin{equation}
\left(\partial_{tt}-3v_{th}^2\nabla^2+\omega_{pe}^2\right)\tilde{n}_{e}=\frac{\omega_{pe}^2}{4\pi e}\nabla^2(\vec{\tilde{A}}_s\cdot\vec{a}_{0}),  \label{as}
\end{equation}
\begin{equation}
\left(\partial_{tt}-c^2\nabla^2+\omega_{pe}^2\right)\vec{\tilde{A}}_s=-4\pi ec^2\tilde{n}_{e}\vec{a}_{0},  \label{ne}
\end{equation}
where $e$ is the charge of electrons, $c$ is the light speed in a vacuum, $\vec{a}_0=\vec{a}_1+\vec{a}_2$ is the amplitude of incident lasers, $v_{th}$ is the electron thermal velocity, $\omega_{pe}$ is the electron plasma frequency, and $\vec{\tilde{A}}_s=\vec{\tilde{A}}_{s1}+\vec{\tilde{A}}_{s2}$ and $\tilde{n}_{e}$ are the perturbations of the scattering light and electron density, respectively. The relation between the normalized amplitude and the laser intensity is $a_0=\sqrt{I_0 (\mathrm{W}/\mathrm{cm}^2)[\lambda(\mu \mathrm{m})]^2/1.37\times 10^{18}}$. Both $\vec{a}_1\cdot\vec{\tilde{A}}_{s2}$ and $\vec{a}_2\cdot\vec{\tilde{A}}_{s1}$ can resonate with the shared Langmuir wave, when the frequency difference $\delta\omega_0=\omega_2-\omega_1>0$ of the double-color beams is smaller enough compared to the SRS growth rate $\Gamma$ \cite{YaoZ2017Effective}. Incident lasers with s-polarization (electric field components perpendicular to the incident plane) are considered in the following discussion for stimulated scattering instabilities, which can cause more intense SRS and SBS than in the p-polarized case (electric field components within the incident plane) \cite{zhao2021mitigation}.

\begin{figure}
\centering
    \begin{tabular}{lc}
        \begin{overpic}[width=1\textwidth]{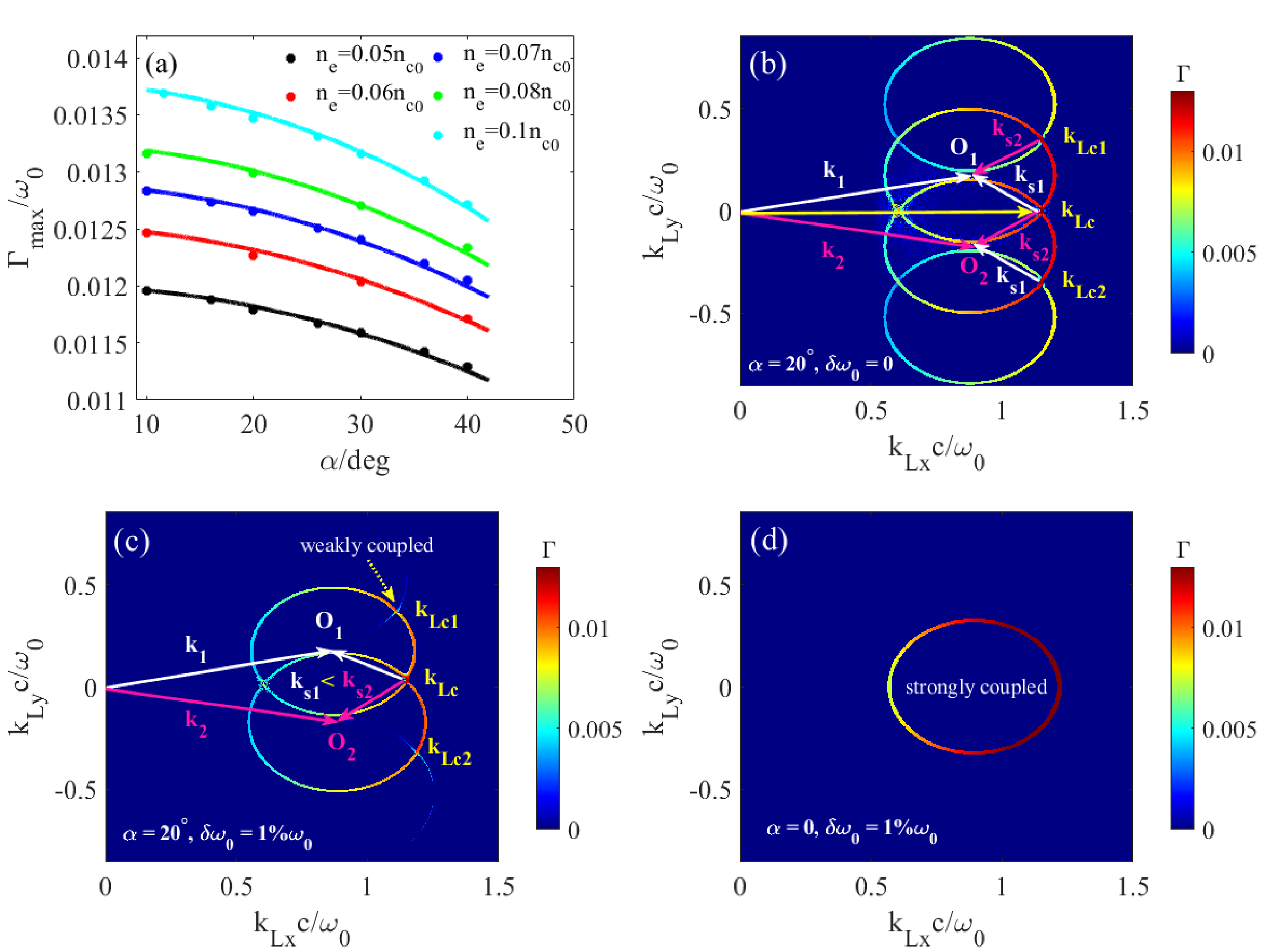}
        \end{overpic}
    \end{tabular}
\caption{(a) The maximum growth rate as a function of the incident angle $\alpha$ under different plasma densities. The markers and continuous lines are the solutions of equations \eqref{ncpl} and \eqref{gth} at $a_1=a_2=0.03$, respectively. (b)-(c) Numerical solutions of SRS dispersion relation (equation \eqref{ncpl}) with $\alpha=20^\circ$ under different frequency differences, where (b) $\delta\omega_0=0$ and (c) $\delta\omega_0=1\%\omega_0$. (d) Numerical solutions of SRS dispersion relation (equation \eqref{cpl0}) with $\delta\omega_0=1\%\omega_0$ and $\alpha=0$. The plasma density is $n_e=0.2n_{c0}$, where $n_{c0}$ is the critical density for the incident laser with frequency $\omega_0$.
    }
\end{figure}

The detail Fourier analysis of equations \eqref{as} and \eqref{ne} is discussed in the Appendix. In the coupling regime $\delta\omega_0\lesssim2\sqrt{2}\Gamma$ \cite{YaoZ2017Effective}, the two beams are collinear if they share the same Langmuir wave and scattering light. The daughter waves of the two beams cannot be distinguished from each other in the phase space, i.e., their instability region are overlapped $\mid\vec{k}_{L1}-\vec{k}_{L2}\mid\lesssim\sqrt{2}(\Delta\kappa_1+\Delta\kappa_2)$, where $\vec{k}_{Lj}$ and $\Delta\kappa_j=a_jk_{Lj}\sqrt{\omega_{pe}(\omega_j-\omega_{pe})/(\omega_j^2-2\omega_j\omega_{pe})}$ are the wavevector of Langmuir wave and the width of instability region driven by the $j$-th beam \cite{zhao2022mitigation}. The widest instability region is found at the backscattering mode. Therefore, the angle of collinear beams should satisfy
\begin{equation}
\alpha<\alpha_c\approx\sqrt{2}\Delta\kappa_0c/\omega_0,
\end{equation}
where $\Delta\kappa_0=a_0k_{L0}\sqrt{\omega_{pe}(\omega_0-\omega_{pe})/(\omega_0^2-2\omega_0\omega_{pe})}$, $\omega_0=(\omega_1+\omega_2)/2$ is the central frequency, $k_{0}c=\sqrt{\omega_0^2-\omega_{pe}^2}$, and $k_{L0}=k_0+\sqrt{\omega_0^2-2\omega_0\omega_{pe}}/c$. Considering for the ICF conditions $a_0\sim0.01$, the critical angle determining the threshold of collinear and non-collinear lights always satisfies $\alpha_c\lesssim4^\circ$.

Actually, if $\alpha<\alpha_c$, then $\mid2\vec{k}_1-2\vec{k}_2\mid\approx\mid\vec{k}_1-\vec{k}_2\mid\sim0$. Thus, the Fourier analysis of equations \eqref{as} and \eqref{ne} lead to
\begin{equation}
\begin{split}
D_p=&\beta\left[a_1^2\left(\frac{1}{D_{s-1}}+\frac{1}{D_{s+1}}\right)+a_2^2\left(\frac{1}{D_{s-2}}+\frac{1}{D_{s+2}}\right)\right.\\
&\left.+a_1a_2\left(\frac{1}{D_{s-1}}+\frac{1}{D_{s+1}}+\frac{1}{D_{s-2}}+\frac{1}{D_{s+2}}\right)\right],
\end{split}
\label{cpl0}
\end{equation}
where $D_{p}=\omega^2-3k^2v_{th}^2-\omega_{pe}^2$, $\beta=\omega_{pe}^2k^2c^2/4$ and $D_{s\pm j}=(\omega\pm\omega_j)^2-(\vec{k}\pm\vec{k}_j)^2c^2-\omega_{pe}^2$ for the $j$-th beam. The above equation \eqref{cpl0} describes the SRS instability driven by a strongly coupled CPL.

Different from the CPL, the scattered lights propagation in different directions are not completely shared by the incident beams \cite{zhao2021mitigation}. The SRS dispersion relation of NCPL can be obtained from the Fourier analysis of equations \eqref{as} and \eqref{ne} at $\alpha>\alpha_c$ as follows
\begin{equation}
\begin{split}
&\left(\frac{D_p}{\beta}-\frac{a_1^2}{D_{s-1}}-\frac{a_2^2}{D_{s-2}}-\frac{a_2^2}{D_{s+2}}-\frac{a_1^2}{D_{s+1}}\right)\left(\frac{D_{p+2-1}}{\beta_{+2-1}}-\frac{a_1^2}{D_{s-1+2-1}}-\frac{a_2^2}{D_{s-1}}-\frac{a_1^2}{D_{s+2}}\right.\\
&\left.-\frac{a_2^2}{D_{s+2+2-1}}\right)\left(\frac{D_{p+1-2}}{\beta_{+1-2}}-\frac{a_1^2}{D_{s-2}}-\frac{a_2^2}{D_{s+1-2-2}}-\frac{a_1^2}{D_{s+1+1-2}}-\frac{a_2^2}{D_{s+1}}\right)\\
&=\left(\frac{a_1a_2}{D_{s-1}}+\frac{a_1a_2}{D_{s+2}}\right)^2\left(\frac{D_{p+1-2}}{\beta_{+1-2}}-\frac{a_1^2}{D_{s-2}}-\frac{a_2^2}{D_{s+1-2-2}}-\frac{a_1^2}{D_{s+1+1-2}}-\frac{a_2^2}{D_{s+1}}\right)\\
&+\left(\frac{a_1a_2}{D_{s-2}}+\frac{a_1a_2}{D_{s+1}}\right)^2\left(\frac{D_{p+2-1}}{\beta_{+2-1}}-\frac{a_1^2}{D_{s-1+2-1}}-\frac{a_2^2}{D_{s-1}}-\frac{a_1^2}{D_{s+2}}-\frac{a_2^2}{D_{s+2+2-1}}\right).
\end{split}
\label{ncpl}
\end{equation}
As an example, we denote $D_{p\pm j\pm n}=(\omega\pm\omega_j\pm\omega_n)^2-3(\vec{k}\pm\vec{k}_j\pm\vec{k}_n)^2v_{th}^2-\omega_{pe}^2$, $\beta_{\pm j\pm n}=\omega_{pe}^2(\vec{k}\pm\vec{k}_j\pm\vec{k}_n)^2c^2/4$, and $D_{s\pm j\pm n}=(\omega\pm\omega_j\pm\omega_n)^2-(\vec{k}\pm\vec{k}_j\pm\vec{k}_n)^2c^2-\omega_{pe}^2$ for the $j$-th and $n$-th beams. By comparing equations \eqref{ncpl} and \eqref{cpl0}, the ponderomotive force of NCPL is reduced to $(2+\sqrt{2})/4\sim85\%$ of that of CPL. According to equation \eqref{ncpl}, the maximum growth rate decreases with an increasing incident angle. Assuming that $\alpha\gg\alpha_c$ and $\omega_1=\omega_2$, the decay rate of $\Gamma_{\mathrm{max}}$ can be estimated as
\begin{equation}
\frac{d\Gamma_{\mathrm{max}}}{d\alpha}\approx-\frac{1}{2}\omega_{pe}(\omega_0-\omega_{pe})\sin2\alpha\left(\frac{1}{\sqrt{\omega_0^2\cos^2\alpha-\omega_{pe}^2}}+\frac{1}{\sqrt{\omega_0^2\cos^2\alpha-2\omega_0\omega_{pe}}}\right).
\label{gth}
\end{equation}
The solution comparisons of equations \eqref{ncpl} and \eqref{gth} are shown in Fig. 1(a), and are well matched in the underdense plasma $n_e\lesssim0.1n_{c0}$. Note that the introduced angle $\alpha\sim\alpha_c$ of the NCPL is sufficient to significantly mitigate LPIs, because the growth rate $\Gamma_{\mathrm{max}}$ decreases very slowly with the angle if $\alpha\gg\alpha_c$. For example, the growth rate satisfies $\Gamma_{\mathrm{max}}(\alpha=40^\circ)/\Gamma_{\mathrm{max}}(\alpha=10^\circ)=94.38\%$ at $n_e=0.05n_{c0}$.

In the decoupling regime $\delta\omega_0>2\sqrt{2}\Gamma$ \cite{YaoZ2017Effective}, the double-color beams develop their own scattering light independently. The SRS dispersion relations of both the CPL and NCPL are reduced to
\begin{equation}
D_p=\beta\left[a_1^2\left(\frac{1}{D_{s-1}}+\frac{1}{D_{s+1}}\right)+a_2^2\left(\frac{1}{D_{s-2}}+\frac{1}{D_{s+2}}\right)\right].   \label{dc}
\end{equation}
A comparison between equations \eqref{cpl0} and \eqref{dc} demonstrates that the ponderomotive force of the decoupled beams is only half that of the strongly coupled beams.

\begin{figure}
\centering
    \begin{tabular}{lc}
        \begin{overpic}[width=1\textwidth]{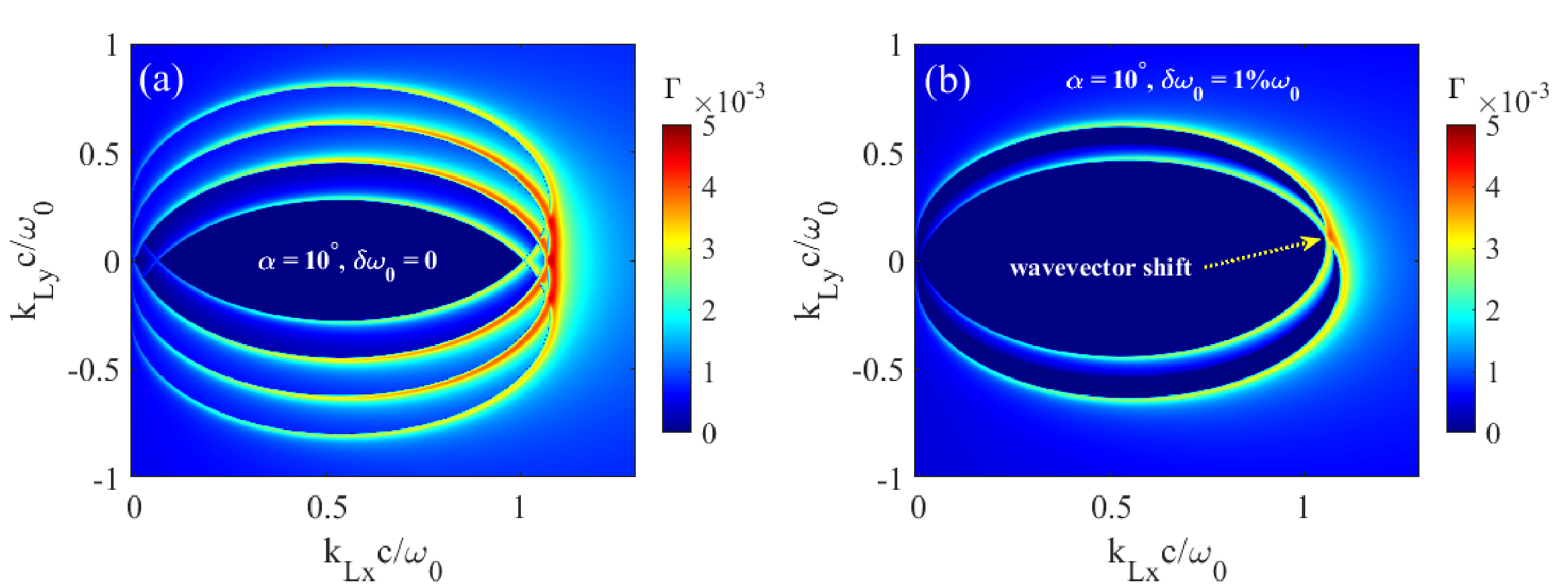}
        \end{overpic}
    \end{tabular}
\caption{Numerical solutions of SBS dispersion relation under different frequency differences (a) $\delta\omega_0=0$ and (b) $\delta\omega_0=1\%\omega_0$. The amplitude and angle of the two incident beams are $a_1=a_2=0.03$ and $\alpha=10^\circ$, respectively. The plasma density and ion mass are $n_e=0.7n_{c0}$ and $m_i=1836m_e$, respectively.
    }
\end{figure}

The numerical solutions of equations \eqref{cpl0} and \eqref{ncpl} are shown in Figs. 1(b)-1(d) for different incident angles and frequency differences. NCPL with a large introduced angle $\alpha=20^\circ$ is considered here for the visualization of the collective behaviors in the phase space. As shown in Fig. 1(b), the scattering light driven by one of the incident lights is shared by the other beam in the coupling regime, and two submodes $\vec{k}_{Lc1}=\vec{k}_1-\vec{k}_{s2}$ and $\vec{k}_{Lc2}=\vec{k}_2-\vec{k}_{s1}$ of the common mode $\vec{k}_{Lc}$ are therefore developed, where $\vec{k}_{sj}$ is the wavevector of the scattering light developed by the $j$-th beam. When a frequency difference $\delta\omega_0=1\%\omega_0$ is introduced into the two beams, the scattered lights are weakly coupled with the sharing modes as exhibited in Fig. 1(c). As shown in Fig. 1(d), the double-colored beams are still strongly coupled for the CPL with $\alpha=0$, even though the frequency difference is $\delta\omega_0=1\%\omega_0$. The above comparisons indicate that the introduced angle can reduce the decoupling threshold of the CPL due to the wavevector mismatch of partial scattering lights.

Actually, equation \eqref{ncpl} can also describe the dispersion relation of SBS with $\beta=\omega_{pi}^2k^2c^2/4$, $\omega_{pi}=\omega_{pe}\sqrt{Zm_e/m_i}$ and $D_p=\omega^2-k^2c_s^2$. Here, $Z$ is the charge number of ions, $c_s$ is the velocity of ion-acoustic wave (IAW), and $m_e$ and $m_i$ are the electron and ion masses, respectively. Similar to SRS, the growth rate of SBS can be reduced by increasing the incident angle. The numerical solutions of SBS dispersion relation with different frequency differences in the near critical density plasma are shown in Fig. 2. According to Fig. 2(a), one finds that many intense submodes are developed by crossing monochromatic beams, where the scattered lights are strongly coupled to the incident beams. However, the scattered lights are no longer shared by the pumps at $\delta\omega_0=1\%\omega_0$ due to the wavevector shift of the shared IAW as shown in Fig. 2(b). Note that an intense CBET resonance between the NCPL beamlets can be driven, if their frequency difference satisfies $\delta\omega_0\sim\mid2\vec{k}_0c_s\sin\alpha\mid$. Therefore, a minimum frequency difference $\delta\omega_0>2k_0c_s\sin\alpha\sim0.2\%\omega_0$ is required for the NCPL to avoid the CBET inside the beamlets. Otherwise, the propagation of NCPL is uncontrolled. We will discuss the optical design for the generation of NCPL with $\delta\omega_0\sim1\%\omega_0$ in the next section. Considering the damping of IAW $\nu_i$, we have $D_p=\omega(\omega+i\nu_i)-k^2c_s^2$. The IAW damping introduces a threshold for development of SBS and CBET, and reduces the linear growth rate.

\begin{figure}
\centering
    \begin{tabular}{lc}
        \begin{overpic}[width=1\textwidth]{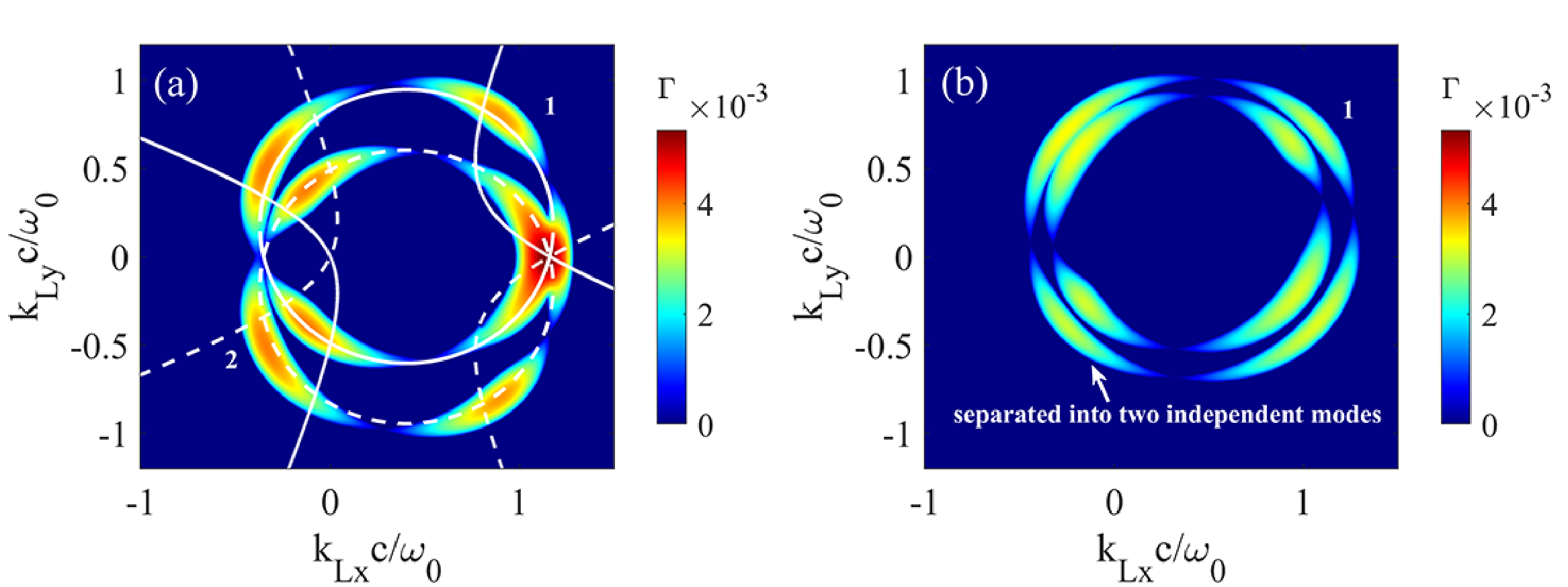}
        \end{overpic}
    \end{tabular}
\caption{Numerical solutions of TPD dispersion relation under different conditions (a) two monochromatic beams with incident angle $\theta=46^\circ$ and $\delta\omega_0=0$, and (b) an NCPL with $\alpha=4^\circ$ and $\delta\omega_0=1\%\omega_0$. The amplitudes of the two incident beams are $a_1=a_2=0.01$ for (a), and the beamlet amplitudes of NCPL are $a_{11}=a_{12}=0.00707$ for (b). The plasma density and electron temperature are $n_e=0.231n_{c0}$ and $T_e=4$keV, respectively. The white lines are the expected wavenumbers for plasmons \cite{vu2010}.
    }
\end{figure}

The fluid equation describing TPD is \cite{kruer1988physics}
\begin{equation}
\left(\partial_{tt}-3v_{th}^2\nabla^2+\omega_{pe}^2\right)\tilde{n}_{e}+c\partial_t(\vec{a}_{0}\cdot\nabla\tilde{n}_{e})-n_0\nabla^2(\vec{a}_{0}\cdot\vec{\tilde{v}}_{fl})=0,
\label{fle-tpd}
\end{equation}
where $n_0$ is the primary density, and $\vec{\tilde{v}}_{fl}$ is the perturbation of electron fluid velocity. Analogous to SRS discussed above, the dispersion relation of TPD driven by NCPL is reduced to
\begin{equation}
\begin{split}
(-\omega^2+\omega_L^2)\mu_{12}\mu_{21}&+\frac{\xi_{12}\mu_{21}(c\omega_{-1}\vec{k}_{-1}\cdot\vec{a}_{0}+k_{-1}^2\gamma)(c\omega\vec{k}\cdot\vec{a}_{0}+k^2\gamma_{-1})}{2}\\
&+\frac{\xi_{21}\mu_{12}(c\omega_{-2}\vec{k}_{-2}\cdot\vec{a}_{0}+k_{-2}^2\gamma)(c\omega\vec{k}\cdot\vec{a}_{0}+k^2\gamma_{-2})}{2}=0,
\end{split}
\label{ds-tpd}
\end{equation}
where $\omega_L=\sqrt{\omega_{pe}^2+3k^2v_{th}^2}$, $\mu_{jn}=2\xi_{jn}(\omega_{-j}^2-\omega_{L-j}^2)-(\omega_{+n-j-j}^2-\omega_{L+n-j-j}^2)(c\omega_{+n-j}\vec{k}_{+n-j}\cdot\vec{a}_0+k_{+n-j}^2\gamma_{-j})(c\omega_{-j}\vec{k}_{-j}\cdot\vec{a}_0+k_{-j}^2\gamma_{+n-j})$,
$\xi_{jn}=2(\omega_{+n-j}^2-\omega_{L+n-j}^2)(\omega_{+n-j-j}^2-\omega_{L+n-j-j}^2)-(c\omega_{+n-j-j}\vec{k}_{+n-j-j}\cdot\vec{a}_0+k_{+n-j-j}^2\gamma_{+n-j})(c\omega_{+n-j}\vec{k}_{+n-j}\cdot\vec{a}_0+k_{+n-j}^2\gamma_{+n-j-j})/2$,
and $\gamma=c\vec{k}\cdot\vec{a}_0\omega/k^2$. Here, $\omega_{L\pm j\pm n}=\sqrt{\omega_{pe}^2+3k_{\pm j\pm n}^2v_{th}^2}$.

The p-polarized lasers are considered for TPD due to the nonzero projection of electric field in the propagation direction. The numerical solutions of equation \eqref{ds-tpd} are displayed in Fig. 3, where the plasma density is $n_e=0.231n_{c0}$, the electron temperature is 4keV, and the amplitudes and incident angle of the two lasers are $a_1=a_2=0.01$ and $\theta=46^\circ$, respectively. The collective TPD modes driven by two incident monochromatic lasers are shown in Fig. 3(a), where the dispersion relation coincides with the intersection of expected wavenumbers for plasmons denoted by white lines \cite{vu2010}. When the monochromatic beam 1 is replaced by the NCPL with beamlet angle 4$^\circ$ and frequency difference $\delta\omega_0=1\%\omega_0$ under the same incident energy, the instability regions are separated into two independent modes with relatively low intensities compared to the monochromatic case, as shown in Fig. 3(b).

According to the above discussions on homogeneous plasmas, the growth rate $\Gamma$ of LPIs is mitigated by the incident angle $\alpha>0$, which demonstrates that the saturation gain $G$ of LPIs in inhomogeneous plasmas will also be reduced \cite{Zhao2019,liu1974raman,rosenbluth1972}. Assuming that the saturation level of CPL is $\delta E_x\exp(G)$, where $\delta E_x$ is the initial perturbations of electrostatic field. If the growth rate $\Gamma$ is reduced to $\Gamma\varepsilon<\Gamma$ by the introduced angle, the saturation level of the convective modes is thus reduced by $1-\exp(\varepsilon^2G)/\exp(G)>1-\exp(0.85^2)/\exp(1)\approx24\%$. The detail discussions on LPIs driven by NCPL in inhomogeneous plasmas are presented in the Appendix.

\begin{figure}
\centering
    \begin{tabular}{lc}
        \begin{overpic}[width=1\textwidth]{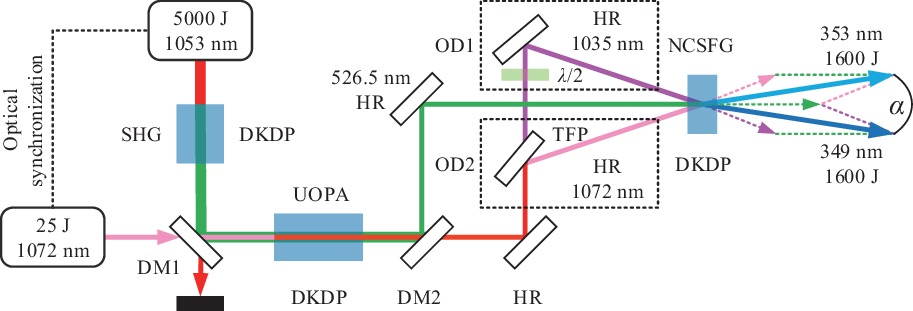}
        \end{overpic}
    \end{tabular}
\caption{Schematic of non-collinear polychromatic light generation. DM, HR, OD, TFP and $\lambda/2$ denote the dichroic mirror, high reflector, optical delay, thin film polarizer, and polarization rotating crystal, respectively.
    }
\end{figure}

\section{Optical design of non-collinear polychromatic driver with double-color beamlets}

In previous CPL scheme, two-color beamlets are generated and amplified via saturated optical parametric amplification, and each monochromatic beamlet is frequency converted independently \cite{zhao2021polychromatic}. Different from that, the two-color beamlets are generated simultaneously within a same DKDP crystal via NCSFG, and the optical schematic of NCPL with detuning $\delta\omega_0\sim1\%\omega_0$ is shown in Fig. 4. The fundamental wave is a spatiotemporal shaping high-energy Nd:glass (1053 nm) laser with a pulse energy and duration of 5 kJ and 3 ns, respectively. A 20-mm-long KD$_2$PO$_4$ (DKDP) crystal is used for SHG (cutting angles of $\Theta = 36.6^\circ$ and $\Phi = 45^\circ$, type-I phase matching). The corresponding energy evolutions of the fundamental wave and second harmonic in the DKDP crystal are displayed in Fig. 5 (a) \cite{Ref01,Ref02,nie2022hybrid,Ref03}. The second harmonic is separated from the residual fundamental wave via dichroic mirror 1 (DM1; 1053 nm and 1072 nm transmission, 526.5 nm reflection) with an output energy of 4.25 kJ, which corresponds to a conversion efficiency of approximately 85\%. Meanwhile, the pre-amplified signal light with a wavelength 1072 nm and energy 25 J is combined with the second-harmonic pump using DM1.

\begin{figure}
\centering
    \begin{tabular}{lc}
        \begin{overpic}[width=1\textwidth]{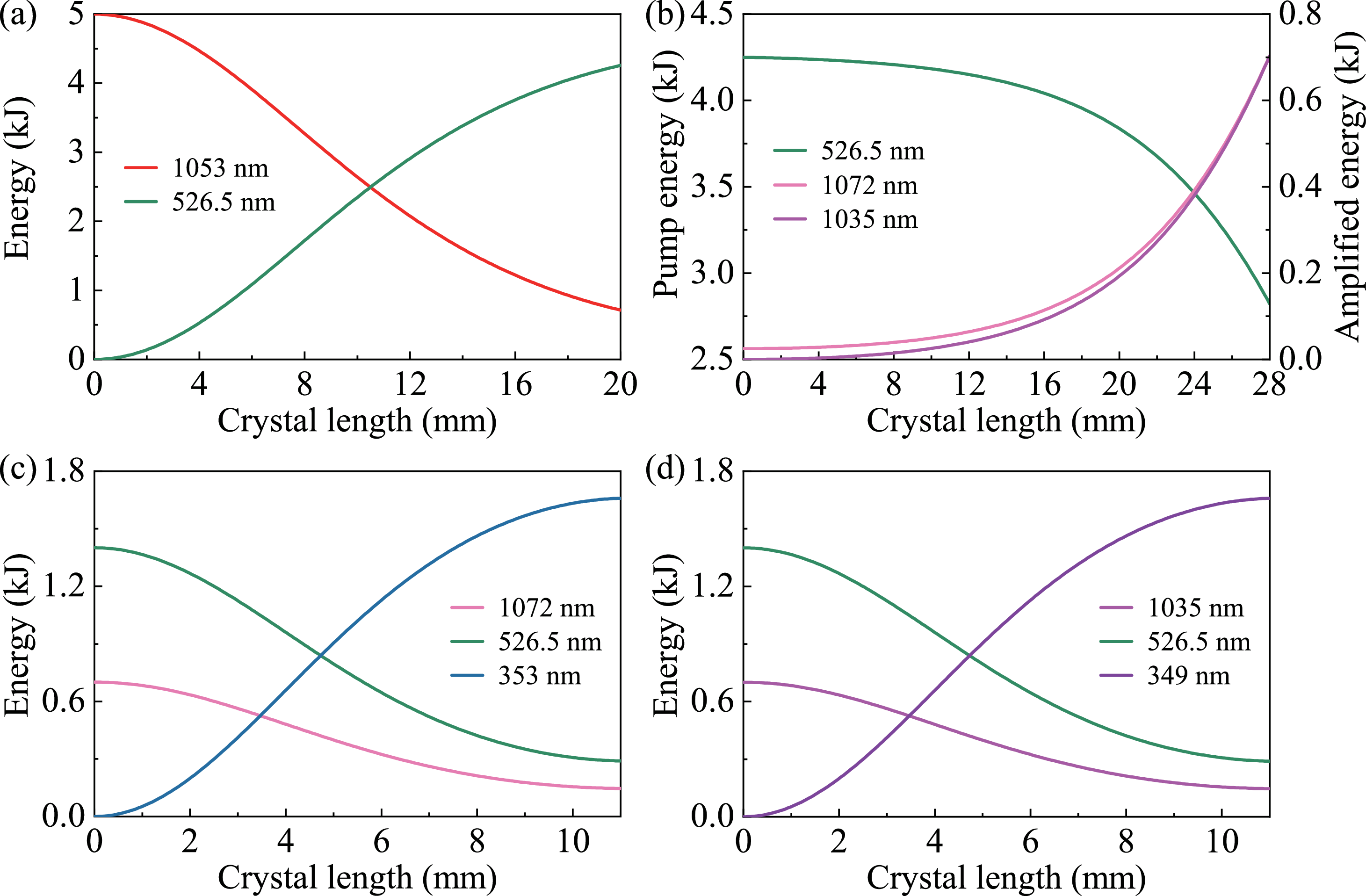}
        \end{overpic}
    \end{tabular}
\caption{Energy evolutions of the interacting waves during the nonlinear processes (a) SHG of fundamental frequency laser, (b) UOPA, (c) and (d) NCSFG of two-color beamlets in a same DKDP.
    }
\end{figure}

Considering the optimal energy ratio required for the subsequent NCSFG, unsaturated amplification is used in the UOPA process. Thus, the 28-mm-long DKDP crystal is employed as an optical parametric amplifier (cutting angles of $\Theta=53.1^\circ$ and $\Phi=0^\circ$, type-II phase matching). The energies of the pump light (526.5 nm), signal light (1072 nm), and generated idler light (1035 nm) as functions of the crystal length are exhibited in Fig. 5(b) \cite{Ref03,Ref05}. After the UOPA, the output energies of the amplified signal light and generated idler light are both 0.7 kJ, and the energy of residual pump light is 2.8 kJ. The polarization directions of the signal light and idler light are vertical due to the type-II phase matching in the UOPA process. Subsequently, the pump light is separated from the signal light and idler light by DM2 (526.5 nm reflection, 1035 nm and 1072 nm transmission), and then the signal light and idler light are separated by the thin film polarizer, which enables the signal light and idler light to be incident at different angles into the NCSFG crystal.

As demonstrated in Fig. 4, a polarization rotating crystal is used to ensure that the polarization of the 1035-nm laser is consistent with that of the 1072-nm laser. An 11-mm-long DKDP crystal (cutting angles of $\Theta=47.9^\circ$ and $\Phi=45^\circ$, type-I phase matching) is used in the NCSFG process to generate two beamlets of 349 nm and 353 nm with a nonzero angle $\alpha$, where the phase-matching angles of the two beamlets are approached \cite{Ref01,Ref06}. The final output energy of each ultraviolet laser is 1.6 kJ, and the total conversion efficiency from the fundamental wave to the ultraviolet NCPL is up to 64\%. The energy evolutions of the generation of 353 nm and 349 nm beamlets are shown in Figs. 5(c) and 5(d), respectively \cite{Ref03,Ref05}. The angle $\alpha$ between the output beamlets can be flexibly adjusted from 0 to $5^\circ$ owing to the adjustable phase-matching angles in the NCSFG. Type-I phase matching is used for the NCSFG, and the corresponding coupling equations of the five-wave mixing are
\begin{align}
\begin{alignedat}{1}
\frac{\partial A_1}{\partial x} &= \frac{i \omega_1}{n_{1o} c} d_{\mathrm{eff}} E_4 E_3^* e^{ i \Delta k_{\mathrm{1}} x}
                                 + \frac{i}{2 k_{1o}} \nabla_\perp^2 A_1, \\
\frac{\partial A_2}{\partial x} &= \frac{i \omega_2}{n_{2o} c} d_{\mathrm{eff}} E_5 E_3^* e^{ i \Delta k_{\mathrm{2}} x}
                                 + \frac{i}{2 k_{2o}} \nabla_\perp^2 A_2, \\
\frac{\partial A_3}{\partial x} &= \frac{i \omega_3}{n_{3o} c} d_{\mathrm{eff}} E_4 E_1^* e^{ i \Delta k_{\mathrm{1}} x}
                                 + \frac{i}{2 k_{3o}} \nabla_\perp^2 A_3
                                 + \frac{i \omega_3}{n_{3o} c} d_{\mathrm{eff}} E_5 E_2^* e^{ i \Delta k_{\mathrm{2}} x}, \\
\frac{\partial A_4}{\partial x} &= \frac{i \omega_4}{n_{4e} c} d_{\mathrm{eff}} E_1 E_3   e^{-i \Delta k_{\mathrm{1}} x}
                                 + \frac{i}{2 k_{4e}} \nabla_\perp^2 A_4
                                 + \tan \rho_{4e} \frac{\partial A_4}{\partial y}, \\
\frac{\partial A_5}{\partial x} &= \frac{i \omega_5}{n_{5e} c} d_{\mathrm{eff}} E_2 E_3   e^{-i \Delta k_{\mathrm{2}} x}
                                 + \frac{i}{2 k_{5e}} \nabla_\perp^2 A_5
                                 + \tan \rho_{5e} \frac{\partial A_5}{\partial y},
\end{alignedat}                                                                                     \label{THG}
\end{align}
where the subscripts 1, 2, 3, 4, and 5 denote the 1035 nm, 1072 nm, 526.5 nm, 349 nm, and 353 nm lasers, respectively. The subscripts $o$ and $e$ represent the ordinary light and extraordinary light, respectively. Here $A$, $n$ and $\rho$ are the complex amplitude of optical field, refractive index and walk-off angle of extraordinary light, respectively. The operator $\nabla_\perp^2=\frac{\partial^2}{\partial y^2}+\frac{\partial^2}{\partial z^2}$, $d_{\mathrm{eff}}$ is the effective nonlinear coefficient, and $\Delta k_\mathrm{1}$ and $\Delta k_{\mathrm{2}}$ are the phase mismatch $\Delta k_\mathrm{1}=k_{4e}-k_{3o}-k_{1o}$ and $\Delta k_\mathrm{2}=k_{5e}-k_{3o}-k_{2o}$, respectively.

Compared to the traditional driver, only the THG of the nonlinear processes is replaced by the NCSFG, which leads to a high efficient and compact NCPL drive system. The UOPA process optimizes the energy conversion of the residual pump light, amplified signal light, and generated idler light in the subsequent NCSFG. Therefore, an extremely low energy loss is observed in the UOPA nonlinear process. Moreover, the NCSFG performed in a single DKDP crystal can significantly reduce the number of crystals and enhance the energy conversion efficiency, resulting in the same efficiency level as the traditional driver scheme.

\section{Simulations for the mitigation of LPIs by non-collinear polychromatic light}

\begin{figure}
\centering
    \begin{tabular}{lc}
        \begin{overpic}[width=1\textwidth]{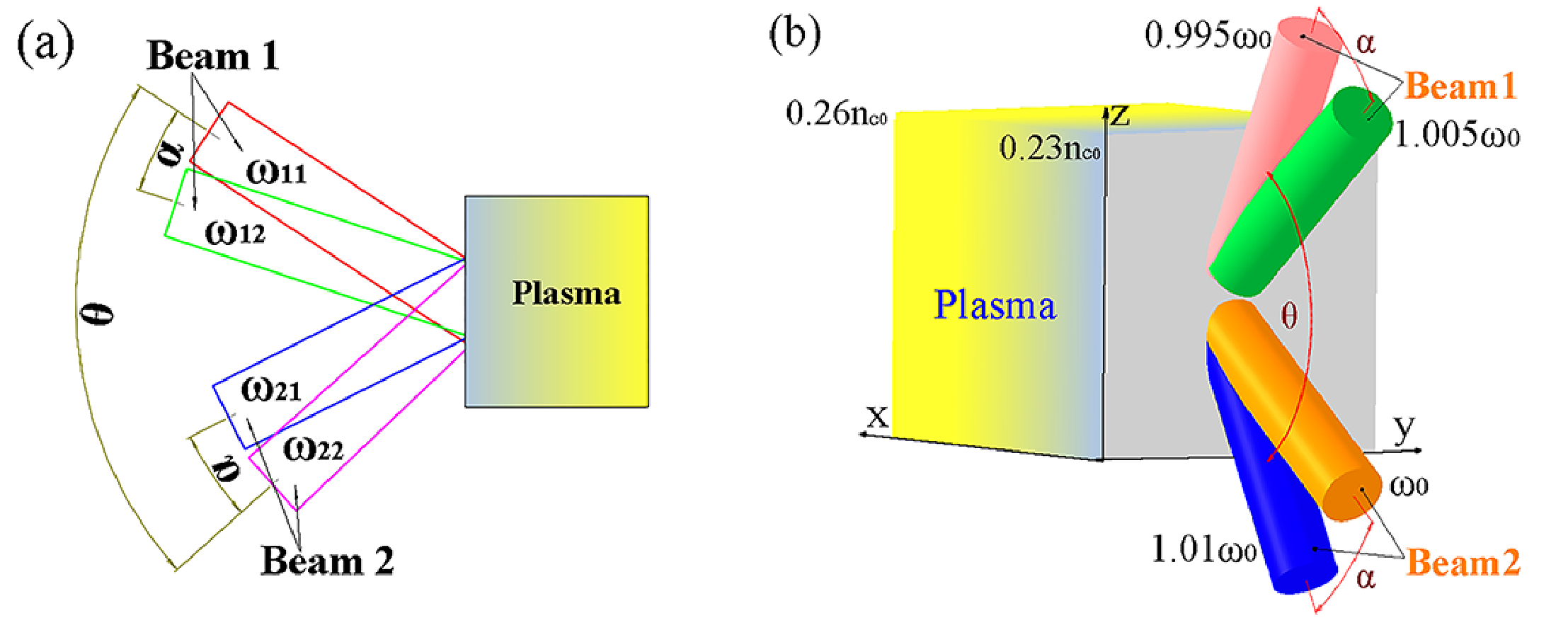}
        \end{overpic}
    \end{tabular}
\caption{Configuration of the incident beams in the (a) 2D simulations and (b) 3D simulations.
    }
\end{figure}

The PIC simulations carried out in this article are use of the OSIRIS code \cite{fonseca2002osiris}. Each monochromatic laser is a third harmonic with intensity of $I_0=1.1\times10^{15}$W/cm$^2$. Under the same incident energy, each beamlet intensity of the polychromatic light is $I_j=5.5\times10^{14}$W/cm$^2$ at a beamlet number of $N=2$. The beamlet angle introduced into the NCPL is denoted by $\alpha$ and the angle between different incident beams is denoted by $\theta$, respectively. The incident lasers are plane wave, and open space boundary conditions are used for the electromagnetic fields. In the following, the spatial axis and time are normalized by the laser wavelength $\lambda=2\pi c/\omega_0$ and period $\tau=\lambda/c$, respectively. The ion mass is $m_i=3672m_e$ with an effective charge number of $Z=1$ for all the simulations. The beamlet angles of each CPL and NCPL are $\alpha=0$ and $\alpha=4^\circ$, respectively. To mitigate the multibeam LPIs \cite{zhao2021polychromatic}, the minimum frequency difference of the beamlets from the adjacent flanges is $\delta\omega_0/2$, as illustrated in Figs. 6(a) and 6(b).

\subsection{Two-dimensional simulation results}

\begin{figure}
\centering
    \begin{tabular}{lc}
        \begin{overpic}[width=1\textwidth]{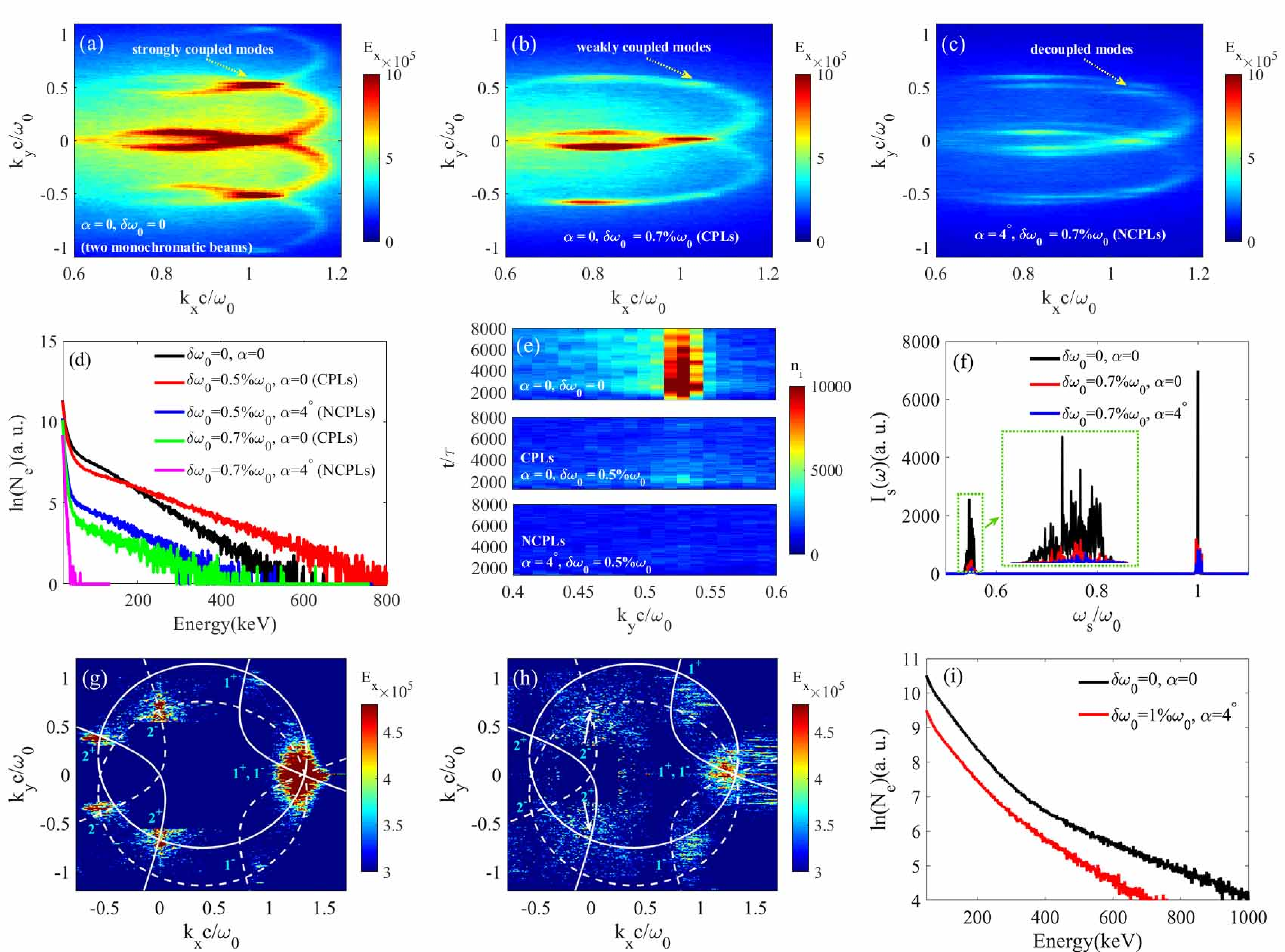}
        \end{overpic}
    \end{tabular}
\caption{(a)-(f) 2D simulations for convective SRS and SBS. Wavenumber distributions of electrostatic field summated from $t=1000\tau$ to $t=8000\tau$ at (a) $\alpha=0$ and $\delta\omega_0=0$ (two monochromatic lasers), (b) $\alpha=0$ and $\delta\omega_0=0.7\%\omega_0$ (two CPLs), and (c) $\alpha=4^\circ$ and $\delta\omega_0=0.7\%\omega_0$ (two NCPLs). (d) Electron energy distributions heated by different types of incident lights at $t=8000\tau$. (e) Transverse wavenumber evolutions of IAW driven by the crossing incident beams under different conditions, where the longitudinal wavenumber is summated from $k_xc=-0.1\omega_0$ to $k_xc=0.1\omega_0$. (f) The frequency spectra of scattering light diagnosed at the transverse center of the left vacuum. The incident angle of the two beams is $\theta=30^\circ$. (g)-(h) 2D simulations for multibeam TPD around 0.237$n_{c0}$ with incident angle $\theta=53^\circ$. Wavenumber distributions of electrostatic field summated from $t=1000\tau$ to $t=3000\tau$ at (g) $\alpha=0$ and $\delta\omega_0=0$ (two monochromatic lasers) and (h) $\alpha=4^\circ$ and $\delta\omega_0=1\%\omega_0$ (two NCPLs). (i) Electron energy distributions heated by different types of incident lights at $t=6000\tau$. The beamlet frequencies $\omega_{11}$, $\omega_{12}$, $\omega_{21}$ and $\omega_{22}$ are $0.9975\omega_0$, $1.0025\omega_0$, $\omega_0$ and $1.005\omega_0$ for the case of $\delta\omega_0=0.5\%\omega_0$, are $0.9965\omega_0$, $1.0035\omega_0$, $\omega_0$ and $1.007\omega_0$ for the case of $\delta\omega_0=0.7\%\omega_0$, and are $0.995\omega_0$, $1.005\omega_0$, $\omega_0$ and $1.01\omega_0$ for the case of $\delta\omega_0=1\%\omega_0$, respectively. The subscript of $\omega_{jn}$ denotes the $n$-th beamlet of the $j$-th incident beam.
    }
\end{figure}

In the 2D simulations for convective SRS and SBS, all the incident lights are s-polarized. The longitudinal plasma density profile is $n_e/n_{c0}=0.19\exp(x/L)$, where $x$ is the longitudinal axis, $L=800\lambda$ is the density scale length, and the density range is $0.19n_{c0}\leq n_e\leq0.21n_{c0}$. The transverse width of the plasma is 80$\lambda$, and the resolutions are $dx/\lambda=dy/\lambda=0.0357$ with 30 particles per cell. The electron and ion temperatures are $T_e=2$ keV and $T_i=0.85$ keV, respectively. The configuration of the incident beams is presented in Fig. 6(a), where the incident angle of the two beams is $\theta=30^\circ$. For the monochromatic lasers, $\omega_{j1}=\omega_{j2}$ and $\alpha=0$ with the subscript denoting the different beamlet of the $j$-th incident beam.

An intense common mode $\vec{k}_{Lc}c=(1,0)\omega_0$ with two submodes $\vec{k}_{Lc1}c=\vec{k}_{1}c-\vec{k}_{s2}c=(1,0.52)\omega_0$ and $\vec{k}_{Lc2}c=\vec{k}_{2}c-\vec{k}_{s1}c=(1,-0.52)\omega_0$ are driven by two monochromatic lasers as shown in Fig. 7(a), where the side-scattered lights are shared by the incident beams. Note that the collective SRS instability is the dominant mode under the multibeam configuration, where the SRS side-scatter of a single beam is not found in the phase plot. For the CPL case, we introduce the frequency difference $\delta\omega_0=\omega_{12}-\omega_{11}=\omega_{22}-\omega_{21}=0.7\%\omega_0$ into each incident beam, but let the double-color beamlets propagation in the same direction, i.e., $\alpha=0$. Compared to the monochromatic lasers, the intensity of Langmuir waves is reduced by the CPLs according to Fig. 7(b). However, the scattering lights are still weakly coupled to the submodes. Under the simulation conditions, the beamlets of each incident beam are non-collinear when the introduced angle satisfies $\alpha\gtrsim3^\circ$. As discussed in the theoretical section, the SRS growth rate of CPL is reduced by 15\% if the double-color beamlets are non-collinear. Therefore, the saturation level $\delta E_x\exp(2)$ of the weakly coupled modes driven by CPL can be reduced by $42.6\%$ with use of NCPL. As illustrated in Fig. 7(c), the two electrostatic modes are decoupled in the phase space at $\alpha=4^\circ$, and the intensity of multibeam SRS is significantly mitigated by the NCPLs.

To further compare the mitigation effects of the CPLs and NCPLs, we perform simulations for an additional case with $\delta\omega_0=0.5\%\omega_0$, where the beamlets of the CPLs are strongly coupled. A comparison of the electron energy distributions heated by different types of lasers is shown in Fig. 7(d). Compared with the monochromatic lasers, one finds that even more electrons are heated to a higher temperature by the strongly coupled CPLs with $\delta\omega_0=0.5\%\omega_0$. However, the hot tail is greatly reduced by the NCPLs with the same frequency difference $\delta\omega_0=0.5\%\omega_0$, which is only slightly higher than the hot tail heated by the CPLs with $\delta\omega_0=0.7\%\omega_0$. The hot electrons are almost completely suppressed by the decoupled NCPLs with $\delta\omega_0=0.7\%\omega_0$. The cross beams transfer their energies through stimulated ion acoustic waves \cite{zhao2021polychromatic,Igumenshchev2012Crossed}, where one of the pump beams is a seed light to develop SBS via beating with the other one. The temporal evolutions of the shared ion modes driven by different types of lights are displayed in Fig. 7(e). An intense CBET mode can be observed for the monochromatic lasers near $\vec{k}_{pi}\approx\vec{k}_1-\vec{k}_2$, that is, $k_{pix}c\approx0$ and $k_{piy}c\approx0.52\omega_0$. However, the instability has not been clearly developed in the case of CPLs with $\delta\omega_0=0.5\%\omega_0$. Note that the intensity of IAW is further reduced by the NCPLs with the same frequency difference as that of the CPLs.

The frequency spectra of scattering light are diagnosed at the transverse center of the left vacuum as shown in Fig. 7(f). An intense convective SRS mode is found at [0.538, 0.555]$\omega_0$ for two monochromatic beams, which indicates that the hot electrons are heated by the collective behavior of multibeam SRS. The intensity of SBS around $\omega_0$ is relatively higher than the convective SRS. However, both the convective SRS and SBS are greatly reduced by the CPLs, and reduced further by use of NCPLs. The significant reduction of scattering light is also found for the other points in the left vacuum. We calculate the reflectivity according to the energy fraction of scattered light propagation into the left vacuum regarding to the incident light. The average reflectivity of monochromatic beams is 8.35\%, which is almost 2.6 times larger than that of NCPLs.

In the 2D simulations for multibeam TPD, the incident lights are p-polarized with incident angle $\theta=53^\circ$. The longitudinal density profile is $n_e/n_{c0}=0.228\exp(x/L)$ with $L=700\lambda$ and the density range of $0.228n_{c0}\leq n_e\leq0.252n_{c0}$. The other parameters are the same as the above simulations for convective SRS and SBS. According to the comparison between Figs. 7(g) and 7(h), the instability regions of TPD are dispersed into a broad range by the decoupled beams, and the intensity of multibeam TPD is significantly inhibited by NCPLs. The temperature and total energy of hot electrons are both reduced by the NCPLs as displayed in Fig. 7(i). The total energy of the hot electrons with kinetic energies $\geqslant$50 keV is reduced by 56.6\% with using polychromatic lights. Therefore, the above simulation results indicate that the decoupling threshold, instability saturation level, and hot electron generation can be significantly mitigated by the NCPLs.

\subsection{Three-dimensional simulation results}

\begin{figure}
\centering
    \begin{tabular}{lc}
        \begin{overpic}[width=1\textwidth]{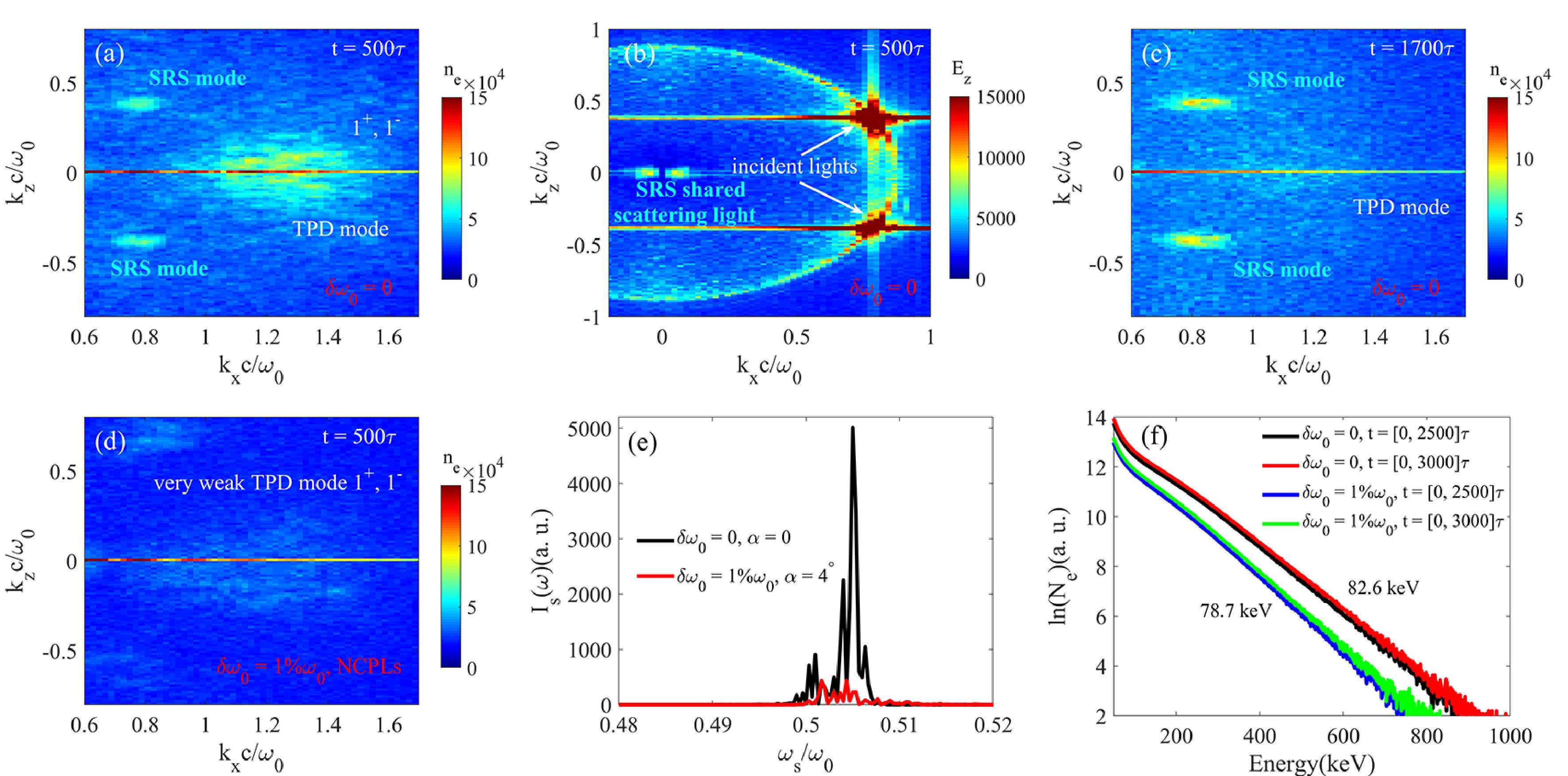}
        \end{overpic}
    \end{tabular}
\caption{$k_x$-$k_z$ wavenumber distributions of electron plasma waves summated from $k_yc/\omega_0=-1$ to $k_yc/\omega_0=1$ at (a) $t=500\tau$ and (c) $t=1700\tau$ for two monochromatic beams, and (d) $t=500\tau$ for two NCPLs, respectively. (b) $k_x$-$k_z$ wavenumber distributions of transverse electric field $E_z$ summated from $k_yc/\omega_0=-1$ to $k_yc/\omega_0=1$ at $t=500\tau$ for two monochromatic beams. (e) The frequency spectra of backscattering light diagnosed at the transverse center of the left vacuum. (f) Time integral energy spectra of hot electrons driven by different types of incident lights in different time windows [0, 2500]$\tau$ and [0, 3000]$\tau$.
    }
\end{figure}

In the 3D simulations, the longitudinal profile of plasma density is $n_e/n_{c0}=0.23\exp(x/L)$ with $L=650\lambda$. The length and height of the plasma are 80$\lambda$ and 74$\lambda$, respectively. The electron and ion temperatures are $T_e=3$ keV and $T_i=1$ keV, respectively. We have set 16 cells per wavelength, and 8 particles per cell. The configuration of the incident beams is presented in Fig. 6(b), where the angle between the two incident beams is $\theta=53^\circ$. As in the 2D simulations, the beamlet angle of each NCPL is $\alpha=4^\circ$. The polarization of incident lights is on the $x-z$ plane, which is perpendicular to the $y-$axis.

Based on Fig. 8(a), the wavenumber distribution of electron plasma waves driven by two monochromatic beams demonstrates that both TPD and SRS have been developed in the early stage of parametric instability excitation, and the intensity of TPD is relatively larger than SRS. Figure 8(b) shows that the two incident beams share a common scattering light with a transverse wavenumber $k_zc=0$. The longitudinal wavenumber of SRS scattering light is $k_xc\sim0$, which indicates that the absolute instability is developed around $n_e\sim0.25n_c$. At $t=1700\tau$, the TPD mode is much weaker than SRS mode according to Fig. 8(c). Different from the 2D simulations for TPD, the polarization of incident laser also has a $z-$projection component, which is perpendicular to the $x-y$ plane in the 3D geometry. Therefore, multibeam SRS sharing a common scattering light is the dominant mode in the later stage. Both SRS and TPD are significantly mitigated by NCPLs at $t=500\tau$, as shown in Fig. 8(d).

The frequency spectra of SRS shared light are diagnosed at the transverse center of the left vacuum as shown in Fig. 8(e). The intensity of common scattering light is significantly reduced by NCPLs. The average reflectivity of backscatter near the transverse center is around 11.69\% for monochromatic beams, which is 4.89 times larger than that of NCPLs. As presented in Fig. 8(f), the time integral energy spectra of hot electrons demonstrate that the electron heating has been saturated at $t=3000\tau$. The temperature of hot electrons with energy larger than 50 keV is about 82.6 keV for the monochromatic beams, which is reduced to 78.7 keV by NCPLs. The hot-electron temperature is only reduced by 4 keV. However, the total energy of the hot electrons with kinetic energies $\geqslant$50 keV is reduced by 57\% with using polychromatic lights. The preheating effect on the implosion is studied using the reduced kinetic model of hot electrons transport in a fluid background \cite{zhao2023Simulations,Ramis2021computer,Delettrez2019Determining}, which indicates that the implosion yield driven by the NCPLs is more than 15 times that of the monochromatic lasers under the direct-drive scheme with megajoule-scale incident energy.

\section{Summary}

In contrast to conventional broadband lasers, we have proposed the LPI model and optical design of NCPL with double-color beamlets propagation in different directions with frequency difference $\gtrsim1\%\omega_0$. Theoretical models of NCPL indicate that the beamlets are non-collinear if the introduced angle satisfies $\alpha\sim4^\circ$ under ICF conditions, and therefore
the wavevector couplings of LPIs are reduced. According to the dispersion relations of stimulated scattering instabilities in homogeneous plasmas, the growth rate of the CPL is reduced by 15\% with using NCPL under the same bandwidth. The saturation level of LPIs driven by CPL in the inhomogeneous plasmas can be reduced by more than 24\% using NCPL with the same bandwidth. Analogous to SRS, the intensities of both SBS and CBET can be further reduced by NCPL. The theoretical model of TPD demonstrates that the compact instability region is dispersed into decoupled submodes with much weak intensity by the NCPL. PIC simulations of convective and absolute instabilities indicate that NCPL with $\delta\omega_0=1\%\omega_0$ is sufficient to effectively mitigate the reflectivity, hot-electron generation, and CBET intensity in laser plasma interactions. Compared to the traditional driver, only the THG of the nonlinear processes is replaced by the NCSFG, which leads to the high efficient and compact NCPL drive system.

\section{Acknowledgement}

The authors appreciate the referees for constructive comments on our manuscript. Y. Z. acknowledge useful discussions with Zhengming Sheng, Anle Lei, Ning Kang and Huiya Liu. This work was supported by the National Natural Science Foundation of China (No. 12005287), the Guangdong Basic and Applied Basic Research Foundation (Grant No. 2023A1515011695), and the Strategic Priority Research Program of Chinese Academy of Sciences (Grant No. XDA25050600). The authors would like to acknowledge the OSIRIS Consortium, consisting of UCLA and IST (Lisbon, Portugal) for providing access to the OSIRIS 4.0 framework.

\section{Appendix}

For a homogeneous plasma, here we mainly discuss the Fourier analysis of equations \eqref{as} and \eqref{ne} as an example. At $\alpha\lesssim\alpha_c$, the coupling modes of $\tilde{n}_{e}(\vec{k}+\vec{k}_j-\vec{k}_n,\omega+\omega_j-\omega_n)$ (denoted by $\tilde{n}_{e+j-n}$) are presented as follows
\begin{equation}
\begin{split}
&\left(\frac{D_{p+j-n}}{\beta_{+j-n}}-\frac{a_n^2}{D_{s-n+j-n}}-\frac{a_j^2}{D_{s-n}}-\frac{a_n^2}{D_{s+j}}-\frac{a_j^2}{D_{s+j+j-n}}\right)\tilde{n}_{e+j-n}\\
&=a_ja_n\left(\frac{\tilde{n}_{e+j-n+j-n}}{D_{s-n+j-n}}+\frac{\tilde{n}_e}{D_{s-n}}+\frac{\tilde{n}_e}{D_{s+j}}+\frac{\tilde{n}_{e+j-n+j-n}}{D_{s+j+j-n}}\right),
\end{split}
\label{n}
\end{equation}
where $j\neq n$. Without loss of generality, we retain only the red-shift terms as follows
\begin{equation}
\begin{split}
&\left(\frac{D_p}{\beta}-\frac{a_1^2}{D_{s-1}}-\frac{a_2^2}{D_{s-2}}\right)\left(\frac{D_{p+2-1}}{\beta_{+2-1}}-\frac{a_1^2+a_1a_2\phi_{21}}{D_{s-1+2-1}}-\frac{a_2^2}{D_{s-1}}\right)\left(\frac{D_{p+1-2}}{\beta_{+1-2}}\right.\\
&\left.-\frac{a_1^2}{D_{s-2}}-\frac{a_2^2+a_1a_2\phi_{12}}{D_{s+1-2-2}}\right)=\left(\frac{a_1a_2}{D_{s-1}}\right)^2\left(\frac{D_{p+1-2}}{\beta_{+1-2}}-\frac{a_1^2}{D_{s-2}}-\frac{a_2^2+a_1a_2\phi_{12}}{D_{s+1-2-2}}\right)\\
&+\left(\frac{a_1a_2}{D_{s-2}}\right)^2\left(\frac{D_{p+2-1}}{\beta_{+2-1}}-\frac{a_1^2+a_1a_2\phi_{21}}{D_{s-1+2-1}}-\frac{a_2^2}{D_{s-1}}\right),
\end{split}
\label{cpl}
\end{equation}
where $\phi_{21}=a_1a_2/(D_{p+2-1+2-1}D_{s+2-1-1}/\beta_{+2-1+2-1}-a_1^2D_{s+2-1-1}/D_{s+2-1+2-1-1}-a_2^2)$ and $\phi_{12}=a_1a_2/(D_{p+1-2+1-2}D_{s+1-2-2}/\beta_{+1-2+1-2}-a_1^2-a_2^2D_{s+1-2-2}/D_{s+1-2+1-2-2})$. If $\alpha<\alpha_c$, then $\mid2\vec{k}_1-2\vec{k}_2\mid\approx\mid\vec{k}_1-\vec{k}_2\mid\sim0$. Thus, the above dispersion relation is reduced to equation \eqref{cpl0}. Otherwise, the wavevector of $\tilde{n}_{e+j-n+j-n}$ is no longer matched with the sharing submodes $\tilde{n}_{e+j-n}$ at $\mid2\vec{k}_1-2\vec{k}_2\mid>\mid\vec{k}_1-\vec{k}_2\mid>0$. The higher-order drive terms $\phi_{12}$ and $\phi_{21}$ of equation \eqref{cpl} are omitted, which reduces the decoupling threshold of different-color beams.

In an inhomogeneous plasma with a density profile $n_e(x)=n_0(1+x/L)$, the double-color beams share the same Langmuir wave $\vec{k}_{Lc}$ at $n_e/n_{cj}<\cos^4\alpha_j/4$ \cite{zhao2021mitigation}, where $n_{cj}$ is the critical density for the $j$-th beam. The wavevectors are always matched in the transverse direction, $k_{jy}=k_{sjy}$. Therefore, the envelop equations of the convective common mode $\vec{k}_{Lc}$ are \cite{Zhao2019,liu1974raman}
\begin{equation}
\nu_s\tilde{A}_s-v_s\partial_x\tilde{A}_s=\frac{\Gamma_\mathrm{max}}{\sqrt{2}}\tilde{n}_{e}\left[\exp\left(i\int K_{1x}dx\right)+\exp\left(i\int K_{2x}dx\right)\right],
\label{in1}
\end{equation}
\begin{equation}
\nu_p\tilde{n}_{e}+v_p\partial_x\tilde{n}_{e}=\frac{\Gamma_\mathrm{max}}{\sqrt{2}}\tilde{A}_s\left[\exp\left(-i\int K_{1x}dx\right)+\exp\left(-i\int K_{2x}dx\right)\right],
\label{in2}
\end{equation}
where $v_s$ and $v_p$ are the group velocities of the scattered light and plasma wave, and $\nu_s$ and $\nu_p$ are the damping of the scattered light and the plasma wave, respectively. Here, $K_{jx}=k_{jx}-k_{sjx}-k_{Lc}$ is the wavenumber mismatch. As an example for beam 1, $K_{1x}=k_{1x}-k_{s1x}-k_{Lc}=(k_{0x}-\delta k_{0x})-(k_{s0x}-\delta k_{s0x})-k_{Lc}=K_{0x}+\delta K_{0x}$, where $\delta K_{0x}=\delta k_{s0x}-\delta k_{0x}=\delta\omega_0[(\omega_0\cos^2\alpha_1-\omega_{pe})/2k_{s0x}c^2-\omega_0\cos^2\alpha_1/2k_{0x}c^2]$, $k_{0x}c=\sqrt{\omega_0^2\cos^2\alpha_1-\omega_{pe}^2}$, and $k_{s0x}c=\sqrt{\omega_0^2\cos^2\alpha_1-2\omega_0\omega_{pe}}$.

In underdense plasmas $n_e\lesssim0.15n_{c0}$, the primary part of the mismatch can be ignored due to $\delta K_{0x}\lesssim0.1\delta\omega_0/c\sim10^{-3}\omega_0/c$. Therefore, its Taylor expansion in the resonance region is $\delta K_{0x}=\delta K'_{0x}x$ with $\delta K'_{0x}=d\delta K_{0x}/dx$. In our previous work \cite{Zhao2019}, we assumed that $\delta K'_{0x}\sim\delta k'_{0x}$ at $n_e<0.1n_{c0}$. The exact expression for the saturation gain in underdense plasmas is \cite{zhao2022mitigation,rosenbluth1972}
\begin{equation}
G=\frac{2\pi\Gamma_{\mathrm{max}}^2}{v_sv_pK'_{0x}}\left[1+\cos\left(\frac{4\nu_p^2\delta K'_{0x}}{K_{0x}'^2v_p^2}\right)\right].
\label{g1}
\end{equation}

In high-density plasmas $n_e\gtrsim0.2n_{c0}$, the primary part of $\delta K_{0x}$ cannot be ignored for beams with a broad bandwidth $\delta\omega_0\sim10^{-2}\omega_0$. The gain coefficient is obtained based on equations \eqref{in1} and \eqref{in2}, as follows
\begin{equation}
G=\frac{2\pi\Gamma_{\mathrm{max}}^2}{v_sv_pK'_{0x}}\left[1+\exp\left(-\frac{4\nu_p\delta K_{0x}}{K'_{0x}v_p}\right)\right].
\label{g2}
\end{equation}

Different from the common mode, the submodes cannot be described by the envelop equations. The wavenumber mismatch of the submodes $\vec{k}_{Lc1}$ and $\vec{k}_{Lc2}$ in the longitudinal direction are $K_{12x}=k_{1x}-k_{s2x}-k_{Lcx}=k_{s1x}-k_{s2x}$ and $K_{21x}=k_{2x}-k_{s1x}-k_{Lcx}=k_{s2x}-k_{s1x}$, respectively. For example, the primary part of $K_{12x}$ is
\begin{equation}
K_{12x0}=-\frac{\delta\omega_0(\omega_0\cos^2\alpha_1-\omega_{pe})}{c\sqrt{\omega_0^2\cos^2\alpha_1-2\omega_0\omega_{pe}}},
\label{k}
\end{equation}
which is much larger than $\delta K_{0x}$ under the same frequency difference $\delta\omega_0$. The submode $\vec{k}_{Lc1}$ is suppressed if the primary part satisfies $\mid K_{12x0}\mid\gtrsim\sqrt{2}\Delta\kappa_{Lc1}$, where the scattering light $\vec{k}_{s2}$ is decoupled from the submode. Therefore, the decoupling condition of submodes can be estimated to
\begin{equation}
\delta\omega_0\gtrsim\frac{a_1k_{Lc1}c}{\omega_0\cos^2\alpha_1-\omega_{pe}}\sqrt{\frac{\omega_{pe}(\omega_0-\omega_{pe})(\omega_0\cos^2\alpha_1-2\omega_{pe})}{\omega_0-2\omega_{pe}}}.
\label{sbc}
\end{equation}
Equation \eqref{sbc} indicates that the decoupling threshold can be reduced by the incident angle particularly in a high-density plasma.

\end{document}